\documentclass[twocolumn,prb]{revtex4}
\usepackage{amssymb}

\input{tcilatex}

\begin{document}

\title{Far-field interaction of focused relativistic electron beams in
electron energy loss spectroscopy of nanoscopic platelets }

\begin{abstract}
A quantum mechanical scattering theory for relativistic, highly focused
electron beams near nanoscopic platelets is presented, revealing a new
excitation mechanism due to the electron wave scattering from the platelet
edges. Radiative electromagnetic excitations within the light cone are shown
to arise, allowed by the breakdown of momentum conservation along the beam
axis in the inelastic scattering process. Calculated for metallic (silver
and gold) and insulating (SiO$_{2}$ and MgO) nanoplatelets, new radiative
features are revealed above the main surface plasmon-polariton peak, and
dramatic enhancements in the electron energy loss probability at gaps of the
'classical' spectra, are found. The corresponding radiation should be
detectable in the vacuum far-field zone, with e-beams exploited as sensitive
'tip-detectors' of electronically excited nanostructures.

PACS number(s): 79.20.Uv, 78.67.Bf, 73.20.Mf, 41.60.-m
\end{abstract}

\author{M. A. Itskovsky$^{1}$, H. Cohen$^{2}$ and T. Maniv$^{1}$}
\date{\today}
\affiliation{$^{1}$Schulich Faculty of Chemistry, Technion-IIT, 32000 Haifa, ISRAEL\\
$^{2}$Weizmann Institute of Science, Chemical Research Support, Rehovot
76100, ISRAEL }
\maketitle

\section{Introduction}

A powerful technique for investigating electromagnetic (EM) field
distribution around nanostructures is provided by very fast (relativistic)
electron beams (e-beams), with typical lateral resolution on an atomic
scale, available in scanning transmission electron microscopes (STEM)\cite%
{Batson83,Howie85,Muller93,Cohen98,Lembrikov03,Cohen03}. As discussed
previously, when the e-beam is restricted to the vacuum near a selected
nanoparticle\cite{Batson83,Howie85}, its EM interaction with surface
plasmons or surface plasmon-polaritons (SPPs) \cite{Raether80} is
reminiscent of the near-field interaction\cite{Cohen98} of subwavelength
optical probes. Several works have recently studied realizations of
Cherenkov radiation excitation within various dielectric media by e-beams
moving in near-field vacuum zones \cite%
{Zabala03,deAbajo98-99,Ochiai04,deAbajo03}. In all the latter works the
energy loss processes were described within a simple classical model in
which the fast electron was assumed to move with a constant velocity along a
straight line trajectory near a finite dielectric medium such that the
energy loss intensity could be obtained from the force exerted on the
electron due to its self-induced electric field through the nearby
dielectric medium. The great simplification achieved by this approach
amounts to reducing the full scattering problem at hand to a problem of
finding the EM field induced by the e-beam in the vacuum around the
dielectric medium. \ The resulting EM field could include Cherenkov-like
radiative components around the e-beam which were restricted, however, to
propagation within the interior of the dielectric medium. \ 

\FRAME{ftbpFU}{3.6264in}{3.0104in}{0pt}{\Qcb{Schematic illustration of the
e-beam configuration with respect to the rectangular platelet used in our
calculations. \ Another configuration, used in Ref.\protect\cite{deAbajo02},
\ where the e-beam is parallel to an edge of the platelet is also shown. }}{%
\Qlb{Fig.1}}{prbfig1.eps}{\special{language "Scientific Word";type
"GRAPHIC";display "USEDEF";valid_file "F";width 3.6264in;height
3.0104in;depth 0pt;original-width 7.5234in;original-height 6.2241in;cropleft
"0";croptop "1";cropright "1";cropbottom "0";filename
'PRBFig1.eps';file-properties "XNPEU";}}

In this paper we present a quantum mechanical theory for the inelastic
scattering of a relativistic highly focused e-beam traveling near
nanoparticles in a `non-touching' aloof configuration\cite{Batson83,Howie85}%
. We show that the electron wave scattering by nanoparticle edges along the
beam axis switches on Cherenkov-like radiation channels which extend into
the vacuum away from the nanoparticle. The resulting far-field coupling
between the electron and the nanoparticle is found to dramatically enhance
various radiative channels in the loss spectrum.

To illustrate our main points we consider here a simple model (see Fig.1)
where the e-beam is propagated in the vacuum along a wide face of a
rectangular nanoplatelet (oriented, e.g., in the $x-y$ plane), and a surface
or guided wave induced by the electron is propagated with a wave number $%
k_{x}$ along the beam axis. The spatially sensitive nature of the
corresponding electron energy loss process arises from the exponential
dependence, $e^{-2K^{\star }b}$, of the EM interaction between the e-beam
and the platelet on the impact parameter $b$. The extinction coefficient, $%
K^{\star }=\sqrt{K^{2}-\left( \omega /c\right) ^{2}}$, with $%
K^{2}=k_{x}^{2}+k_{y}^{2}$ , determines the tail of the evanescent field in
the vacuum for values of $k$\ outside the light-cone, i.e. for $K>\omega /c$%
. Inside the light-cone, i.e. for $K<\omega /c$ , $K^{\star }$ is purely
imaginary and the corresponding interaction becomes spatially oscillating,
allowing the electron to exchange photons with the particle far away into
the vacuum. This striking mechanism has been overlooked in the recent
literature of STEM-electron energy loss spectroscopy (EELS), since the
excitation by an electron moving in the vacuum with a classical velocity $v$%
, has been restricted to a constant longitudinal wavenumber $k_{x}=\omega
/v>\omega /c$, implying EM coupling to the nanoparticle which is restricted
to the evanescent tail near the surface.

Our model calculation is applied to two types of nanoscopic platelets,
conducting platelets made of silver or gold, and dielectric platelets made
of insulators such as silica or magnesia. \ For both types of nanoparticles
we find significant loss signals in the low energy range of the spectrum,
where the electron-hole excitation probability is either zero (for the
insulator) or very small (for the metals), exhibiting far-field (radiative)
characteristics. In particular, specific SPP modes of the silver platelet,
which penetrate into the light cone, can be excited by the external e-beam,
leading to new features in the EEL spectrum which decay weakly with the
beam-platelet distance.

\section{Model and formulation}

Following Ref.\cite{Cohen98}, the focused e-beam is described here as a
one-dimensional wave, propagating along the $x$-axis, while in the
transverse ($y-z$) directions it is described by a wave-packet localized
within a smoothly converging cross section along the beam axis, whose shape
is assumed to be squared for the sake of simplicity. The corresponding
Green's function for the noninteracting focused e-beam may be therefore
written in the general form: 
\begin{eqnarray}
G_{e}^{\left( 0\right) }\left( \overrightarrow{r},\overrightarrow{r}^{\prime
};t\right) &=&\frac{1}{2L}\dsum\limits_{p_{x}}e^{iq_{x}\left( x-x^{\prime
}\right) }\dsum\limits_{\overrightarrow{q}_{tr}}e^{i\varepsilon
_{p_{x},p_{tr}}t/\hbar } \\
&&\chi _{\overrightarrow{q}_{tr}}\left( y,z;x\right) \chi _{\overrightarrow{q%
}_{tr}}^{\ast }\left( y^{\prime },z^{\prime };x^{\prime }\right)  \nonumber
\end{eqnarray}%
where $p_{x}=\hbar q_{x}$ is the longitudinal (along the beam axis) electron
momentum, $\varepsilon _{p_{x},p_{tr}}=\sqrt{%
p_{x}^{2}c^{2}+m_{0}^{2}c^{4}+c^{2}p_{tr}^{2}}$ its total relativistic
energy eigenvalue, with $m_{0}$ the electron rest mass, \ $\overrightarrow{p}%
_{tr}=\hbar \overrightarrow{q}_{tr}$ $\ $($\ \overrightarrow{q}_{tr}=\left(
q_{y},q_{z}\right) $ ) its transverse momentum, and $e^{iq_{x}x}\chi _{%
\overrightarrow{q}_{tr}}\left( y,z;x\right) $ the corresponding e-beam
eigenfunction (see Appendix A).

The electromagnetic (EM) interaction between the e-beam and the platelet may
be described effectively by the Hamiltonian:

\begin{equation}
\widehat{H}_{EM}\left( \overrightarrow{r},\left\{ s\right\} \right) \approx
-e\Phi \left( \overrightarrow{r},\left\{ s\right\} \right) -e\frac{\widehat{p%
}_{x}}{mc}A_{x}\left( \overrightarrow{r},\left\{ s\right\} \right)
\label{Hem}
\end{equation}%
where $\Phi $ and $A_{x}$ are the scalar and $x$-component of the EM
four-vector potential respectively, $\overrightarrow{r}$ is the electron
position vector, $\left\{ s\right\} $ is a collective symbol for the
position vectors of the platelet charges, and $m=m_{0}/\sqrt{1-\left(
v/c\right) ^{2}}$ the dynamic electron mass.

To first order of the perturbation theory with respect to the EM interaction
Hamiltonian, $\widehat{H}_{EM}$ , the probability for the e-beam to go,
during the time interval $\tau $, from initial to final eigen states when
the platelet initial state is the ground state is:

\begin{eqnarray}
&&\sum_{\alpha _{f}}\left\vert K_{e\left( i\rightarrow f\right) ,\alpha
_{0}\rightarrow \alpha _{f}}^{\left( 1\right) }\left( \tau \right)
\right\vert ^{2} \\
&=&\left( \frac{1}{\hbar }\right) ^{2}\sum_{\alpha _{f}}\left\vert \Delta
\left( \varepsilon _{p_{x}^{i},\overrightarrow{q}_{tr}^{i}}+\varepsilon
_{\alpha _{0}},\varepsilon _{p_{x}^{f},\overrightarrow{q}_{tr}^{f}}+%
\varepsilon _{\alpha _{f}};\tau \right) \right\vert ^{2}  \nonumber \\
&&\frac{1}{2L}\int_{-L}^{L}dx^{\prime }\int dy^{\prime }\int dz^{\prime
}\chi _{\overrightarrow{q}_{tr}^{f}}\left( y^{\prime },z^{\prime };x^{\prime
}\right) \chi _{\overrightarrow{q}_{tr}^{i}}^{\ast }\left( y^{\prime
},z^{\prime };x^{\prime }\right)  \nonumber \\
&&e^{-iq_{x}^{i}x^{\prime }}\left\langle \alpha _{0}\left\vert \widehat{H}%
_{EM}\left( \overrightarrow{r}^{\prime }\right) \right\vert \alpha
_{f}\right\rangle e^{iq_{x}^{f}x^{\prime }}  \nonumber \\
&&\frac{1}{2L}\int_{-L}^{L}dx\int dy\int dz\chi _{^{\overrightarrow{q}%
_{tr}^{f}}}^{\ast }\left( y,z;x\right) \chi _{\overrightarrow{q}%
_{tr}^{i}}\left( y,z;x\right)  \nonumber \\
&&e^{-iq_{x}^{f}x}\left\langle \alpha _{f}\left\vert \widehat{H}_{EM}\left( 
\overrightarrow{r}\right) \right\vert \alpha _{0}\right\rangle
e^{iq_{x}^{i}x}  \nonumber
\end{eqnarray}%
where the the sum is over the platelet final states $\alpha _{f}$ , and : $%
\Delta \left( \varepsilon ,\varepsilon ^{\prime };\tau \right) \equiv \frac{%
\exp \left[ i\tau \left( \varepsilon -\varepsilon ^{\prime }\right) /\hbar %
\right] -1}{\left[ i\left( \varepsilon -\varepsilon ^{\prime }\right) /\hbar %
\right] }$\ . \ In the limit when: \ $\tau \rightarrow \infty $ , \ $%
\left\vert \Delta \left( \varepsilon ,\varepsilon ^{\prime };\tau \right)
\right\vert ^{2}\rightarrow 2\pi \hslash \tau \delta \left( \varepsilon
-\varepsilon ^{\prime }\right) =2\tau \func{Re}\int_{0}^{\infty }dt\exp %
\left[ it\left( \varepsilon -\varepsilon ^{\prime }\right) /\hbar \right] $
, and so the rate of change of scattering probability of the e-beam $%
R_{e\left( i\rightarrow f\right) }\equiv \frac{d}{d\tau }\sum_{\alpha
_{f}}\left\vert K_{e\left( i\rightarrow f\right) ,\alpha _{0}\rightarrow
\alpha _{f}}^{\left( 1\right) }\left( \tau \right) \right\vert ^{2}$ , \ \ $%
\tau \rightarrow \infty $:

is given by:

\begin{eqnarray}
&&R_{e\left( i\rightarrow f\right) }=\frac{4\pi }{\hbar }\sum_{\alpha _{f}}
\\
&&\func{Re}\left\{ 
\begin{array}{c}
\int_{0}^{\infty }dt\exp \left[ it\left( \varepsilon _{p_{x}^{i},%
\overrightarrow{q}_{tr}^{i}}+\varepsilon _{\alpha _{0}}-\varepsilon
_{p_{x}^{f},\overrightarrow{q}_{tr}^{f}}-\varepsilon _{\alpha _{f}}\right)
/\hbar \right] \\ 
\frac{1}{2L}\int_{-L}^{L}dx^{\prime }\frac{1}{2L}\int_{-L}^{L}dx \\ 
\int dy^{\prime }\int dz^{\prime }\chi _{\overrightarrow{q}_{tr}^{f}}\left(
y^{\prime },z^{\prime };x^{\prime }\right) \chi _{\overrightarrow{q}%
_{tr}^{i}}^{\ast }\left( y^{\prime },z^{\prime };x^{\prime }\right) \\ 
\int dy\int dz\chi _{\overrightarrow{q}_{tr}^{f}}^{\ast }\left( y,z;x\right)
\chi _{\overrightarrow{q}_{tr}^{i}}\left( y,z;x\right) \\ 
\left\langle \alpha _{0}\left\vert e^{-iq_{x}^{i}x^{\prime }}\widehat{H}%
_{EM}\left( \overrightarrow{r}^{\prime }\right) e^{iq_{x}^{f}x^{\prime
}}\right\vert \alpha _{f}\right\rangle \\ 
\left\langle \alpha _{f}\left\vert e^{-iq_{x}^{f}x}\widehat{H}_{EM}\left( 
\overrightarrow{r}\right) e^{iq_{x}^{i}x}\right\vert \alpha _{0}\right\rangle%
\end{array}%
\right\}  \nonumber
\end{eqnarray}%
where the inclusion of all terms under the real part symbol is justified by
the reality of the total expression written in the last five rows within the
curly brackets. \ 

Using the relations: $\ e^{-iq_{x}^{f}x}\widehat{H}_{EM}\left( 
\overrightarrow{r}\right) e^{iq_{x}^{i}x}=e^{i\left(
q_{x}^{i}-q_{x}^{f}\right) x}\widehat{H}_{EM}^{p_{x}^{f}}\left( 
\overrightarrow{r}\right) $, with: $\widehat{H}_{EM}^{p_{x}^{f}}\left( 
\overrightarrow{r}\right) \equiv \left( -e\right) \left[ \widehat{\Phi }%
\left( \overrightarrow{r}\right) -\frac{p_{x}^{f}}{mc}\widehat{A}_{x}\left( 
\overrightarrow{r}\right) \right] $, \ the rate of change of probability for
the scattering of the e-beam can be rewritten in the form:

\begin{eqnarray}
&&R_{e\left( i\rightarrow f\right) }=\sum_{\overrightarrow{q}_{tr}^{i},%
\overrightarrow{q}_{tr}^{f}}e^{-\beta \frac{\left( \hbar q_{tr}^{i}\right)
^{2}}{2m_{0}}}\times  \label{RateProb1} \\
&&\frac{4\pi }{\hbar }\func{Re}\left\{ 
\begin{array}{c}
\frac{1}{2L}\int_{-L}^{L}dx^{\prime }e^{-i\Delta q_{x}x^{\prime }}\frac{1}{2L%
}\int_{-L}^{L}dxe^{i\Delta q_{x}x} \\ 
\int dy^{\prime }\int dz^{\prime }\chi _{\overrightarrow{q}_{tr}^{f}}\left(
y^{\prime },z^{\prime };x^{\prime }\right) \chi _{\overrightarrow{q}%
_{tr}^{i}}^{\ast }\left( y^{\prime },z^{\prime };x^{\prime }\right) \\ 
\int dy\int dz\chi _{\overrightarrow{q}_{tr}^{f}}^{\ast }\left( y,z;x\right)
\chi _{\overrightarrow{q}_{tr}^{i}}\left( y,z;x\right) \\ 
\int_{0}^{\infty }dte^{i\omega t}\left\langle \widehat{H}_{EM}^{p_{x}^{i}}%
\left( x^{\prime },y^{\prime },z^{\prime };t\right) \widehat{H}%
_{EM}^{p_{x}^{f}}\left( x,y,z;0\right) \right\rangle%
\end{array}%
\right\}  \nonumber
\end{eqnarray}%
where 
\begin{equation}
\Delta q_{x}\equiv \left( q_{x}^{i}-q_{x}^{f}\right) \approx \left( \omega
/v\right) +\hbar \left[ \left( q_{tr}^{f}\right) ^{2}-\left(
q_{tr}^{i}\right) ^{2}\right] /2mv  \label{Deltaq_x}
\end{equation}%
is the longitudinal momentum transfer of the e-beam (see Appendix A),\ $%
\overrightarrow{q}_{tr}^{i}$ and $\overrightarrow{q}_{tr}^{f}$ the e-beam
asymptotic transverse momenta, initial and final respectively, and $\hbar
\omega \equiv \left( \varepsilon _{p_{x}^{i},\overrightarrow{q}%
_{tr}^{i}}-\varepsilon _{p_{x}^{f},\overrightarrow{q}_{tr}^{i}}\right) $ its
energy loss.\ Note that the width $\beta ^{-1}$of the Gaussian distribution
function, is introduced in Eq.(\ref{RateProb1}) to account for the high
transverse-energy cutoff caused to the e-beam by the objective aperture. It
is related to the length $L$ of the region around the beam focal plane used
in our model as a normalization factor for the electron wave functions.

The interaction potential, $H_{EM}\left( x,y,z\right) $ , between the
platelet and an external electron at $\left( x,y,z\right) $ is nearly
independent of $x$ for $\left\vert x\right\vert \ll a^{\star }$, and decays
to zero at least as quickly as $1/x^{2}$ for $\left\vert x\right\vert
>a^{\star }$ (see, e.g., Ref.\cite{Ferrell87}). \ Under these circumstances
the limits of the integrations over $x$ and $x^{\prime }$ in the above
expression may be set at $-a^{\ast }$ and $a^{\ast }$, rather than at $-L$
and $L$. \ The correlation function $\left\langle \widehat{H}%
_{EM}^{p_{x}^{i}}\left( x^{\prime },y^{\prime },z^{\prime };t\right) 
\widehat{H}_{EM}^{p_{x}^{f}}\left( x,y,z;0\right) \right\rangle $ can be
expressed in terms of the relevant components of the 4-tensor photon Green's
function $D_{\nu ,\mu }\left( \overrightarrow{r}^{\prime },\overrightarrow{r}%
;t\right) $, $\nu ,\mu =0,1,2,3$ ($\leftrightarrow ct,x,y,z$), as:%
\begin{eqnarray*}
&&\left\langle \widehat{H}_{EM}^{p_{x}^{i}}\left( x^{\prime },y^{\prime
},z^{\prime };t\right) \widehat{H}_{EM}^{p_{x}^{f}}\left( x,y,z;0\right)
\right\rangle \\
&=&i\left[ 
\begin{array}{c}
D_{0,0}\left( \overrightarrow{r}^{\prime },\overrightarrow{r};t\right) +%
\frac{p_{x}^{i}}{mc}D_{0,1}\left( \overrightarrow{r}^{\prime },%
\overrightarrow{r};t\right) \\ 
+\frac{p_{x}^{f}}{mc}D_{1,0}\left( \overrightarrow{r}^{\prime },%
\overrightarrow{r};t\right) +\frac{p_{x}^{f}p_{x}^{i}}{\left( mc\right) ^{2}}%
D_{1,1}\left( \overrightarrow{r}^{\prime },\overrightarrow{r};t\right)%
\end{array}%
\right] \text{ , } \\
\text{\ }t &>&0
\end{eqnarray*}

For the sake of simplicity we may assume translational invariance of the
platelet dielectric properties in the $x-y$ plane, that is take: $D_{\nu
,\mu }\left( \overrightarrow{r}^{\prime },\overrightarrow{r};t\right)
=D_{\nu ,\mu }\left( x-x^{\prime },y-y^{\prime },z^{\prime },z;t\right) $.\
For an impact parameter $b$ smaller than the platelet sides along the $x$
and $y$ axes (i.e. $b\ll 2a^{\star },2b^{\star }$ ) this assumption may be
justified, though it is inconsistent with the breakdown of momentum
conservation in the beam-platelet scattering event considered here (see a
more detailed discussion below).

Substituting into the above expression for $R_{e\left( i\rightarrow f\right)
}$ and rearranging the integrations we find that:%
\begin{eqnarray*}
R_{e\left( i\rightarrow f\right) } &=&-\frac{4\pi e^{2}}{\hbar }\sum_{%
\overrightarrow{q}_{tr}^{i},\overrightarrow{q}_{tr}^{f}}e^{-\beta \frac{%
\left( \hbar q_{tr}^{i}\right) ^{2}}{2m_{0}}}\func{Im}\{\int dk_{x}\int
dk_{y} \\
&&%
\begin{array}{c}
\frac{1}{2L}\int_{-a^{\ast }}^{a^{\ast }}dx^{\prime }e^{-i\left( \Delta
q_{x}-k_{x}\right) x^{\prime }} \\ 
\int dz^{\prime }\int dy^{\prime }e^{ik_{y}y^{\prime }}\chi _{%
\overrightarrow{q}_{tr}^{f}}\left( y^{\prime },z^{\prime };x^{\prime
}\right) \chi _{\overrightarrow{q}_{tr}^{i}}^{\ast }\left( y^{\prime
},z^{\prime };x^{\prime }\right) \\ 
\frac{1}{2L}\int_{-a^{\ast }}^{a^{\ast }}dxe^{i\left( \Delta
q_{x}-k_{x}\right) x} \\ 
\int dz\int dye^{-ik_{y}y}\chi _{\overrightarrow{q}_{tr}^{f}}^{\ast }\left(
y,z;x\right) \chi _{\overrightarrow{q}_{tr}^{i}}\left( y,z;x\right) \\ 
D^{p_{x}^{f},p_{x}^{i}}\left( k_{x},k_{y},\omega ;z^{\prime },z\right) \}%
\end{array}%
\end{eqnarray*}

where $D^{p_{x}^{f},p_{x}^{i}}\left( k_{x},k_{y},\omega ;z^{\prime
},z\right) $

$=\int_{0}^{\infty }dte^{it\omega }D^{p_{x}^{f},p_{x}^{i}}\left(
k_{x},k_{y};z^{\prime },z;t\right) $, 
\begin{eqnarray}
&&D^{p_{x}^{f},p_{x}^{i}}\left( k_{x},k_{y};z^{\prime },z;t\right) = \\
&&%
\begin{array}{c}
D_{0,0}\left( k_{x},k_{y};z^{\prime },z;t\right) +\frac{p_{x}^{i}}{mc}%
D_{0,1}\left( k_{x},k_{y};z^{\prime },z;t\right) + \\ 
\frac{p_{x}^{f}}{mc}D_{1,0}\left( k_{x},k_{y};z^{\prime },z;t\right) +\frac{%
p_{x}^{f}p_{x}^{i}}{\left( mc\right) ^{2}}D_{1,1}\left(
k_{x},k_{y};z^{\prime },z;t\right)%
\end{array}
\nonumber
\end{eqnarray}%
and $D_{\nu ,\mu }\left( k_{x},k_{y};z^{\prime },z;t\right) $ is the spatial
Fourier transform of $D_{\nu ,\mu }\left( x-x^{\prime },y-y^{\prime
},z^{\prime },z;t\right) $ with wavevector $\overrightarrow{K}=\left(
k_{x},k_{y}\right) $.

\ 

Now, the 4-tensor photon propagator in the vacuum (i.e. at $z,z^{\prime
}\leq 0$ ) has the form \cite{Maniv82}: 
\begin{eqnarray*}
&&D_{\nu ,\mu }\left( k_{x},k_{y},\omega ;z^{\prime },z\right) \\
&=&\frac{\eta _{\nu }}{2\pi K^{\star }}\left[ \delta _{\nu ,\mu
}e^{-K^{\star }\left\vert z^{\prime }-z\right\vert }-r_{\nu ,\mu }\left(
k_{x},k_{y},\omega \right) e^{K^{\star }\left( z^{\prime }+z\right) }\right]
\end{eqnarray*}%
in which the relevant part is associated only with the second term within
the square brackets (i.e. that associated with the image potential of the
e-beam). Using this expression and recalling that $1/\left\vert K^{\star
}\right\vert $ is typically much larger than the beam transverse dimension,
so that the extreme confinement of the e-beam wave functions $\chi _{%
\overrightarrow{q}_{tr}^{i,f}}$ under the integrals over $z$ and $z^{\prime
} $ restrict their values to a narrow region near $z^{\prime }=z=-b$ ,we
have:%
\begin{eqnarray*}
&&R_{e\left( i\rightarrow f\right) }\approx \frac{4\pi e^{2}}{\hbar }\sum_{%
\overrightarrow{q}_{tr}^{i},\overrightarrow{q}_{tr}^{f}}e^{-\beta \frac{%
\left( \hbar q_{tr}^{i}\right) ^{2}}{2m_{0}}} \\
&&\times \func{Im}\left\{ 
\begin{array}{c}
\int dk_{x}\int dk_{y}\frac{e^{-2K^{\star }b}}{2\pi K^{\star }}r^{f,i}\left(
k_{x},k_{y},\omega \right) \\ 
\frac{1}{2L}\int_{-a^{\ast }}^{a^{\ast }}dx^{\prime }e^{-i\left( \Delta
q_{x}-k_{x}\right) x^{\prime }} \\ 
J^{\ast }\left( \overrightarrow{q}_{tr}^{f},q_{tr}^{i};k_{y},K^{\star
};x^{\prime }\right) \\ 
\frac{1}{2L}\int_{-a^{\ast }}^{a^{\ast }}dxe^{i\left( \Delta
q_{x}-k_{x}\right) x} \\ 
J\left( \overrightarrow{q}_{tr}^{f},q_{tr}^{i};k_{y},K^{\star };x\right)%
\end{array}%
\right\}
\end{eqnarray*}%
where: \ 
\begin{equation}
\begin{array}{c}
r^{f,i}\left( k_{x},k_{y},\omega \right) =r_{0,0}\left( k_{x},k_{y},\omega
\right) +\frac{\hbar q_{x}^{i}}{mc}r_{0,1}\left( k_{x},k_{y},\omega \right) +
\\ 
\frac{\hbar q_{x}^{f}}{mc}r_{1,0}\left( k_{x},k_{y},\omega \right) +\frac{%
\hbar ^{2}q_{x}^{f}q_{x}^{i}}{\left( mc\right) ^{2}}r_{1,1}\left(
k_{x},k_{y},\omega \right)%
\end{array}
\label{r^if}
\end{equation}

and: 
\begin{eqnarray}
&&J\left( \overrightarrow{q}_{tr}^{f},q_{tr}^{i};k_{y},K^{\star };x\right)
\label{J(if)} \\
&\equiv &\int dz\int dye^{-ik_{y}y}\chi _{\overrightarrow{q}_{tr}^{f}}^{\ast
}\left( y,z;x\right) \chi _{q_{tr}^{i}}\left( y,z;x\right)  \nonumber
\end{eqnarray}

Finally, denoting: 
\begin{eqnarray}
&&I\left( \overrightarrow{q}_{tr}^{f};\overrightarrow{q}_{tr}^{i};k_{y},K^{%
\star };\left( \Delta q_{x}-k_{x}\right) \right)  \label{Kinematical} \\
&\equiv &\frac{1}{2L}\int_{-a^{\ast }}^{a^{\ast }}dxe^{i\left( \Delta
q_{x}-k_{x}\right) x}J\left( \overrightarrow{q}%
_{tr}^{f},q_{tr}^{i};k_{y},K^{\star };x\right)  \nonumber
\end{eqnarray}

the scattering rate is rewritten as:

\begin{eqnarray}
&&R_{e\left( i\rightarrow f\right) }=\frac{2e^{2}}{\hbar }\int dk_{x}\int
dk_{y}  \label{RateProbab} \\
&&\func{Im}\left[ \frac{r^{f,i}\left( k_{x},k_{y},\omega \right) }{K^{\star }%
}e^{-2K^{\star }b}\right]  \nonumber \\
&&\times \sum_{\overrightarrow{q}_{tr}^{i},\overrightarrow{q}%
_{tr}^{f}}e^{-\beta \frac{\left( \hbar q_{tr}^{i}\right) ^{2}}{2m_{0}}%
}\left\vert I\left( \overrightarrow{q}_{tr}^{f};\overrightarrow{q}%
_{tr}^{i};k_{y},K^{\star };\left( \Delta q_{x}-k_{x}\right) \right)
\right\vert ^{2}  \nonumber
\end{eqnarray}

\bigskip

\section{The 'classical' approximation and beyond}

The theory developed in the previous section can be further simplified
without losing its main physical content by employing several
approximations. \ In the long wavelengths limit discussed in Ref.\cite%
{Maniv82} we find that (see Appendix B):

\begin{eqnarray}
&&\func{Im}\left[ e^{-2K^{\ast }b}r\left( \overrightarrow{k},\omega \right)
/K^{\star }\right]  \label{DielectResponse} \\
&\approx &\func{Im}\left\{ \left[ \left( K^{\star }/k^{2}\right)
f_{e}+\left( \left( v/c\right) ^{2}-\left( \omega /ck\right) ^{2}\right)
f_{o}/K^{\ast }\right] e^{-2K^{\ast }b}\right\} ,  \nonumber
\end{eqnarray}%
where 
\[
f_{e}=\left( \varepsilon ^{2}K^{\ast 2}-Q^{2}\right) /D_{e}^{+}D_{e}^{-},\
f_{o}=\left( K^{\ast 2}-Q^{2}\right) /D_{o}^{+}D_{o}^{-} 
\]%
\ 
\begin{eqnarray*}
D_{e}^{+} &=&\varepsilon K^{\ast }+Q\tanh \left( Qc^{\star }\right)
,D_{e}^{-}=\varepsilon K^{\ast }+Q\coth \left( Qc^{\star }\right) \\
D_{o}^{+} &=&K^{\ast }+Q\tanh \left( Qc^{\star }\right) ,D_{o}^{-}=K^{\ast
}+Q\coth \left( Qc^{\star }\right)
\end{eqnarray*}%
$Q\ =\sqrt{K^{2}-(\omega /c)^{2}\varepsilon \left( \omega \right) }$, and $%
\varepsilon \left( \omega \right) $ is the local bulk dielectric function of
the platelet. In the limit of a semi-infinite medium the resulting
expression reduces (see Appendix B) to the surface dielectric response
function obtained in Ref.\cite{Wang96} by using Maxwell's equations with
macroscopic boundary conditions.

The standard classical approximation for the loss function \cite{Wang96} is
obtained from Eq.(\ref{RateProbab}) by making the following assumptions: (1)
the e-beam transverse momentum distribution function $J\left( 
\overrightarrow{q}_{tr}^{f},\overrightarrow{q}_{tr}^{i};k_{y},K^{\star
};x\right) $ is a constant, that is equivalent to a $\delta $-function in
the corresponding real-space transverse coordinates, (2) the contribution of
the transverse energy to the longitudinal momentum transfer $\Delta q_{x}$
(see Eq.(\ref{Deltaq_x})) can be neglected, and (3) the effective particle
size, $a^{\star }$, appearing as an integration limit along the beam axis,
is infinite. Assumption (3), in conjunction with (1), yields the
conservation of longitudinal momentum, i.e. $\Delta q_{x}-k_{x}=0$, which
together with assumption (2) imposes the fixed condition $k_{x}=\left(
\omega /v\right) $.

It is interesting to note that usually assumption (2) is not strictly
satisfied since the contribution of the transverse energy to $\Delta q_{x}$: 
$\hbar \left[ \left( q_{tr}^{f}\right) ^{2}-\left( q_{tr}^{i}\right) ^{2}%
\right] /2mv\approx q_{tr}\Delta q_{tr}/\left( mv/\hbar \right) \sim \pm
q_{tr}^{2}/q_{x}^{i}$ , can be as large in magnitude as $\left( \omega
/v\right) $. \ As an example, at $\hbar \omega \sim 10$ eV, $\left( \omega
/v\right) $ $\sim 0.05$ nm$^{-1}$, whereas the transverse beam-wavenumber
uncertainty, $\left\vert \Delta q_{tr}\right\vert \sim q_{tr}\sim 2\pi /l$
(with a typical value of $l\sim 0.6$ nm for the beam radius) is $10$ nm$%
^{-1} $, so that for $\varepsilon _{^{i}}=100$ keV , where $q_{x}^{i}\sim
1500$ nm$^{-1}$, $q_{tr}^{2}/q_{x}^{i}\sim 0.07$ nm$^{-1}$.

In the present paper we focus on the most interesting violation of the
'classical' approximation outlined above, allowing $a^{\star }$ to be a
finite length, which reflects an effective range of the actual beam-particle
interaction along the beam axis. \ Consequently the longitudinal momentum
distribution around $\Delta q_{x}-k_{x}=0$, defined by the integral in Eq.(%
\ref{Kinematical}), is smeared and many wavenumbers $k_{x}$ inside the
light-cone start contributing to the loss rate, Eq.(\ref{RateProbab}).

The condition for the smearing to be significant is $\pi /a^{\star }\gtrsim
\left( \omega /v\right) $, so that typically for frequencies $\omega $ in
the visible range, $a^{\star }$ should be smaller than $200$ nm.
Nanoplatelets of those lengths should dramatically enhance radiative
excitations by the e-beam, previously overlooked in the literature; see e.g.
Ref.\cite{RitchieHowie88} where it was argued that recoil effects in STEM
should be negligible for valence electron excitations. Recoil is only a
classical remnant of the present effect and of less general appearance. In
particular, it vanishes for large media, such as the porous film
investigated in Ref. \cite{Zabala03}, for which (if made sufficiently thin)
the quantum mechanical momentum uncertainty along the e-beam axis remains
significant.

It should be stressed that, for the sake of simplicity, the platelet
dielectric response is calculated by assuming its wide faces to be
infinite.\ A fully consistent treatment of the breakdown of translation
invariance is expected, however, to further enhance all radiative channels.
The calculation of the kinematical factor $I\left( \overrightarrow{q}%
_{tr}^{f};\overrightarrow{q}_{tr}^{i};k_{y},K^{\star };\left( \Delta
q_{x}-k_{x}\right) \right) $, responsible for the longitudinal momentum
uncertainty in our model, from the integral in Eq.(\ref{Kinematical}), could
generate artificial oscillations by the sharp cutoff of the integral at $%
x=\pm a^{\star }$. To avoid such oscillations we use an equivalent Gaussian
distribution function in our actual calculations. The corresponding smooth
cutoff is, in fact, more realistic than that appearing in Eq.(\ref%
{Kinematical}) since it arises from the attenuation of the e-beam-platelet
interaction at $\left\vert x\right\vert $ values larger than $a^{\ast }$. \
In any event, the exact form of the corresponding distribution function is
of no great importance for the main purpose of our present paper.

\bigskip

\section{Results and Discussions}

\subsection{Silver and Gold Nanoplatelets}

As a first example we calculate the EEL function of a $100$ nm long silver
and gold platelets for an external $100$ keV e-beam at various impact
parameters (see Figs.2 and 3). To analyze the various SPP resonances one may
consider the zeros of the denominator of the extraordinary wave amplitude $%
f_{e}$ in Eq.(\ref{DielectResponse}) in the complex $K$-plane. With the
experimental optical dielectric function, $\varepsilon (\omega )$, for
silver \cite{Palik91} the resulting dispersion relation (inset, Fig.(2))
exhibits a rather flat branch of $\omega (\func{Re}K)$ inside the
light-cone, which can be attributed to radiative SPP, seen as a mirror image
of the usual non-radiative SP dispersion curve with respect to the
light-line. The sector of $\omega (\func{Re}K)$ connecting the two branches
across the light line has a vanishing negative slope, where $\func{Im}%
K(\omega )\propto \func{Im}\varepsilon (\omega )$ has a sharp peak. The
sharp dip in the EEL spectrum just above the classical SP frequency (at $3.8$
eV) reflects these closely related features.

\FRAME{ftbpFU}{3.4765in}{2.879in}{0pt}{\Qcb{EEL spectra (solid lines)\ of a $%
100$ keV e-beam propagating parallel to the $x$-axis of a rectangular Ag
platelet (with half sides: $a^{\star }=50$ nm along $x$ , and $c^{\star }=10$
nm along $z$ , see Fig.1) at impact parameters $b=10,20,40$ nm above its
wide $x-y$ face. The experimental optical dielectric function, $\protect%
\varepsilon (\protect\omega )$, for silver \protect\cite{Palik91} has been
exploited. \ Dashed lines represent spectra calculated by the classical
theory. Inset: surface plsmon polariton (SPP) dispersion curves, $\protect%
\omega \left( \func{Re}K\right) ,$ $\protect\omega \left( \func{Im}K\right) $
in the complex $K$-plane for silver. \ The indicated values of $K$ and $%
\protect\omega $ are normalized by\ \ $K_{n}=\protect\omega _{n}/c,$ and $%
\protect\omega _{n}=10$ $eV$, respectively. }}{\Qlb{Fig.2}}{prbfig2.eps}{%
\special{language "Scientific Word";type "GRAPHIC";maintain-aspect-ratio
TRUE;display "USEDEF";valid_file "F";width 3.4765in;height 2.879in;depth
0pt;original-width 7.5234in;original-height 6.2241in;cropleft "0";croptop
"1";cropright "1";cropbottom "0";filename 'PRBFig2.eps';file-properties
"XNPEU";}}At slightly higher frequencies the EEL signals exhibit a
pronounced rise due to the enhanced SPP density of states associated with
the flat radiative SPP branch. These peculiar features are missing in the
loss spectra of the gold platelet, shown in Fig.3.

\FRAME{ftbpFU}{3.4852in}{2.8858in}{0pt}{\Qcb{The same as Fig.2 for a
platelet made of gold (Au). Note the absence of the sharp dips appearing
just above the main plamon peaks in the corresponding Ag spectra (Fig.2). }}{%
\Qlb{Fig.3}}{prbfig3.eps}{\special{language "Scientific Word";type
"GRAPHIC";maintain-aspect-ratio TRUE;display "USEDEF";valid_file "F";width
3.4852in;height 2.8858in;depth 0pt;original-width 7.5234in;original-height
6.2241in;cropleft "0";croptop "1";cropright "1";cropbottom "0";filename
'PRBFig3.eps';file-properties "XNPEU";}}The EEL intensity in this spectral
region exhibits attenuation with increasing impact parameter significantly
weaker than the corresponding attenuation of the main SP peak calculated in
the classical limit. The radiative nature of the beam-particle coupling
shown in Figs.(2,3) is even more pronounced in the low energy region below
the main SP peak, where the classically calculated signal drops to very
small values. Here our calculated EEL function exhibits a pronounced broad
band with linearly increasing intensity for increasing frequency and almost
no attenuation with increasing impact parameter. These features are due to
the fact that the loss signal well below the main SP frequency is dominated
by the contribution from the ordinary wave amplitude $f_{o}$, appearing in
Eq.(\ref{DielectResponse}), which is singularly enhanced near the light line
(where $K^{\ast }\rightarrow 0$ ), and thus reflecting the nearly pure
(transverse) photonic nature of the excitations by the e-beam in this
'classically forbidden' region.

The results of our calculations may be compared to the experimental data
reported in Ref.\cite{Bosman07} for silver and gold nano rods and
ellipsoids.\ Fig.4 shows our calculated EEL spectra for three silver
platelets with $c^{\star }=15$ nm and $a^{\star }=10,15,30$ nm at impact
parameter $b=10$ nm. \ The shown curves may be compared to the spectrum in
Ref.\cite{Bosman07} obtained for a silver ellipsoid with a long half-axis%
\emph{\ }($\sim 30$\ nm ) and two short half-axes ($\sim 15$\ nm)\emph{\ }at
impact parameter $\sim 10$ nm above the ellipsoid wide face.

\FRAME{ftbpFU}{3.5683in}{2.9547in}{0pt}{\Qcb{The same as Fig.2 for three Ag
platelets with half sides along the beam axis $a^{\star }=10,15,30$ nm, and
half width $c^{\star }=15$ nm, and with the e-beam at an impact parameter $%
b=10$ nm. The dashed line represents the corresponding classical spectrum
(i.e. for $a^{\star }\rightarrow \infty $). }}{\Qlb{Fig.4}}{prbfig4.eps}{%
\special{language "Scientific Word";type "GRAPHIC";maintain-aspect-ratio
TRUE;display "USEDEF";valid_file "F";width 3.5683in;height 2.9547in;depth
0pt;original-width 7.5234in;original-height 6.2241in;cropleft "0";croptop
"1";cropright "1";cropbottom "0";filename 'PRBFig4.eps';file-properties
"XNPEU";}}The two lower curves, particularly those corresponding to $%
a^{\star }=10$ nm, exhibit good agreement with the relevant experimental
data. Specifically, in addition to the very good agreement of the calculated
main plasmon peak position ($\simeq 3.45$ eV) with the experimental one, the
intensities ratio ($\sim 2$) between the main plasmon peak and the high
energy broad peak, and the extent and magnitude of the low energy tail shown
in Fig.4, are seen to agree pretty well with the corresponding experimental
results. In contrast a large intensities ratio ($\sim 8$) and a very small
low energy tail characterize the classical curve shown in Fig.4, both
indicating the importance of the quantum effects predicted by our theory.
Note that an important feature of our calculated spectra, the large dip just
above the main plasmon peak, which is missing in the experimental data, is
shown to develop only at relatively large values of $a^{\star }$ (i.e. for $%
a^{\star }>20$ nm ).

\bigskip

\subsection{Insulating Nanoplatelets}

The situation in the forbidden energy gap region of semiconductors and
insulators is in a sense an extreme case of the effect demonstrated in the
low energy region of Fig.2: The EEL spectra shown in Fig.5 are calculated
for an external $100$ keV e-beam, propagating parallel to the $x-y$ face of
a $100$ nm long SiO$_{2}$ platelet with half thickness $c^{\ast }=50$\ nm,
at different impact parameters $b$. The spectra reveal a pronounced
double-peak structure within the forbidden gap region, which does not decay
with increasing $b$ values. Strictly speaking, this structure reduces to a
single broad peak for platelets of widths $c^{\ast }\lesssim 10$\ nm,
reflecting a finite-size effect. Similarly to the situation with the silver
and gold platelets well below the main SP peak, the strong radiative nature
of this feature arises from the ordinary wave amplitude $f_{o\text{ }}$,
corresponding to the excitation of purely transverse EM waves, polarized
within the $x-y$ plane, which totally dominates the loss signal in the
forbidden gap region.

\FRAME{ftbpFU}{3.7101in}{3.0718in}{0pt}{\Qcb{EEL spectra (solid lines) of a $%
100$ keV e-beam propagating parallel to the $x$-axis of a rectangular SiO$%
_{2}$ platelet at distances $b=2$ and $8$ nm above its wide ($x-y$) face. \
The platelet half-sides along the $x$, and $z$ axes are: $a^{\star }=50$ nm,
and $c^{\star }=50$ nm respectively. The corresponding spectra (dashed
lines) obtained from the classical theory are also shown for comparison. \
Note the close similarity of the classical spectrum for $b=8$ nm with the
one obtained in Ref.\protect\cite{deAbajo02} for the same e-beam velocity at
nearly the same impact parameter parallel to a sharp wedge (see Fig.1). }}{%
\Qlb{Fig.5}}{prbfig5.eps}{\special{language "Scientific Word";type
"GRAPHIC";display "USEDEF";valid_file "F";width 3.7101in;height
3.0718in;depth 0pt;original-width 7.5234in;original-height 6.2241in;cropleft
"0";croptop "1";cropright "1";cropbottom "0";filename
'PRBFig5.eps';file-properties "XNPEU";}}

The spectra shown in Fig.5 may be compared to the results reported in Ref.%
\cite{deAbajo02} for an electron moving parallel to a $90^{\circ }$ SiO$_{2}$
wedge at a distance of $8.5$ nm (see Fig.(1)). The pronounced radiative
broad band within the gap region, obtained in our calculation, dramatically
contrasts the vanishing loss signal shown there in\emph{\ }Fig.(4) for an
electron beam with the same velocity ( $v=0.54c$) and nearly the same impact
parameter. The lack of far-field coupling in the latter theoretical approach
restricted the fast external e-beam to excitation of EM waves confined
within the dielectric medium \cite{Zabala03}, similar to ordinary waveguide
modes which can develop within a thin SiO$_{2}$ slab in the forbidden gap
region where $\func{Re}\varepsilon \left( \omega \right) \approx 2$ , and $%
\func{Im}\varepsilon \left( \omega \right) \rightarrow 0$. \ For an ideal
planar geometry (as assumed in our calculation of the dielectric response
function $r\left( \overrightarrow{K},\omega \right) $), the corresponding
waveguide modes appear as extremely narrow resonances which can not be
excited by an e-beam with $\Delta q_{x}$ values outside the light cone due
to the vanishingly small dielectric damping, $\func{Im}\varepsilon \left(
\omega \right) $.

Such radiation excitations become possible for the nonplanar geometries
studied in Refs.\cite{Zabala03},\cite{deAbajo02} even under the rigid e-beam
trajectory approximation exploited there (but only above a threshold beam
energy considerably higher than $100$ keV) due to the translational
symmetry-broken dielectric media considered in their calculations. Yet, the
corresponding Cherenkov-like channels remain fundamentally different from
the ones we propose: \ The opening of scattering channels with wave numbers
inside the light cone allows coupling of the e-beam to the \textit{continuum}
of EM modes which are extended into the vacuum perpendicular to the platelet
wide face. The relative strength of the present radiative mechanism may be
further appreciated by noting the calculated spectra near a sharp SiO$_{2}$
wedge in Ref.\cite{deAbajo02},where in spite of the geometrical enhancement
of near-field Cherenkov coupling, beam energies far above $100$ keV were
needed there for `switching on' such channels.

Finally, it is instructive to compare our predicted loss spectrum of an
external $100$ keV e-beam propagating above a MgO platelet with half-sides $%
a^{\star }=c^{\star }=50$ nm at an impact parameter $b=2$ nm (see Fig.6) to
the experimental data reported in Ref.\cite{Aizpurua97} for a MgO smoke cube
of $100$ nm size. The overall agreement is good, including the occurrence,
in both the calculated spectrum and the experimental data, of a broad,
nonvanishing signal within the forbidden gap region, which is missing in the
classically calculated spectrum. In this gap region the

\FRAME{ftbpFU}{3.6434in}{3.0173in}{0pt}{\Qcb{EEL spectra (solid lines) of a $%
100$ keV e-beam propagating parallel to the $x$-axis of rectangular MgO
platelets at a distance $b=2$ nm above their wide ($x-y$) faces. \ The
platelets half sides along the $x$ and $z$ axes are $a^{\star }=50$ nm, and $%
c^{\star }=50,100$ nm, respectively. \ The corresponding classical ($%
a^{\star }\rightarrow \infty $) results (dashed lines) are also shown. Note
the finite size oscillations of the calculated loss signal inside the
forbidden energy gap with a period roughly proportianal to $1/c^{\star }$. }%
}{}{prbfig6.eps}{\special{language "Scientific Word";type
"GRAPHIC";maintain-aspect-ratio TRUE;display "USEDEF";valid_file "F";width
3.6434in;height 3.0173in;depth 0pt;original-width 7.5234in;original-height
6.2241in;cropleft "0";croptop "1";cropright "1";cropbottom "0";filename
'PRBFig6.eps';file-properties "XNPEU";}}calculated spectrum exhibits a
smooth oscillatory structure associated with the multiple reflection of the
generated radiation between the two parallel faces of the platelet
perpendicular to the z-axis. This finite-size effect is peculiar to the
far-field radiative modes found in the present paper for platelets confined
in the direction along the e-beam axis,\emph{\ }and is different from
(though related to) the extremely sharp resonances associated with the
waveguide modes developed in an 'ideal' (i.e. wide laterally) planar
dielectric thin film. Thus, the classical approach applied to such an
'ideal' film yields usually (i.e. except for extremely rare coincidences of
the loss energy with the resonant frequencies) null loss intensity, whereas
in\emph{\ }our quantum calculations the continuous window of wavenumbers
inside the light-cone removes the stringent resonant conditions and allows
the appearance of a significant loss intensity in the entire gap region. It
is interesting to note that the average calculated signal inside the gap
region increases smoothly with increasing frequency from zero up to nearly
the interband threshold where its intensity relative to the loss main peaks
(at $\sim $ $14$ and $20$ eV) is about $1/6$. \ This ratio is remarkably
close to the corresponding relative intensity observed experimentally in Ref.%
\cite{Aizpurua97}.

\section{Conclusion}

Applying a quantum-mechanical approach to the scattering problem of highly
focused relativistic e-beams near nanoplatelets, we have shown that
Cherenkov-like radiation of STEM e-beams, discussed recently in the
literature \cite{Zabala03,deAbajo98-99,Ochiai04}, has a much broader scope
than originally presented. Dramatic enhancements of radiative channels arise
from the breakdown of momentum conservation along the e-beam axis in the
inelastic process due to scattering of the electron wave by the nanoparticle
edges. Further enhancements, realized due to the extreme lateral confinement
of the e-beam and its associated transverse momentum uncertainty\cite%
{Cohen98}, have not been considered in detail here. The radiation predicted
to be emitted from both conducting and insulating nanoplatelets can be
generated at impact parameters larger than the evanescent tail of the
excited surface EM modes due to the oscillatory distance dependence of the
electron-platelet interaction for momentum transfers within the light-cone.
\ Consequently, this radiation should have a significant propagation
component perpendicular to its main direction along the e-beam axis.

Large deviations from the classical EEL signal are found to persist also at
small impact parameters, which can be readily tested experimentally. The
results of our calculations for silver platelets seem to agree pretty well
with the experimental data reported in Ref.\cite{Bosman07} for silver nano
ellipsoids. Furthermore, experimental observation of loss signals within the
forbidden energy gap of MgO cubes of $100$ nm size by Aizpurua et al. \cite%
{Aizpurua97} seems as well to support our main prediction.

\section{Appendix A}

In this appendix we specialize our general model of the focused e-beam to
allow a more detailed discussion of some aspects of EELS experiments in STEM
pertinent to the subject under study in this paper. \ We employ the
relativistic Schrodinger's wave equation:

\begin{eqnarray*}
\left[ -\left( \frac{\partial ^{2}}{\partial y^{2}}+\frac{\partial ^{2}}{%
\partial z^{2}}\right) -\frac{\partial ^{2}}{\partial x^{2}}\right] \psi
\left( x,y,z\right) &=&\widetilde{\varepsilon }^{2}\psi \left( x,y,z\right) 
\text{ , } \\
\text{\ }\widetilde{\varepsilon }^{2} &\equiv &\varepsilon ^{2}/\hslash
^{2}c^{2}-m_{0}^{2}c^{2}/\hbar ^{2}
\end{eqnarray*}%
subject to the boundary conditions:\ 

\begin{eqnarray}
\psi \left( x,y,z\right) &=&0\ ,\ \text{for}\ \ l\left( x\right) \geq y\geq 0%
\text{ }  \TCItag{A1}  \label{BoundCond} \\
\text{ \ \ \ \ and for \ }l\left( x\right) -b &\geq &z\geq -b\text{ } 
\nonumber \\
\text{with }\text{: } &&l\left( x\right) =l_{0}+\alpha \left\vert
x\right\vert \text{ \ , \ }\alpha \ll 1  \nonumber
\end{eqnarray}

Due to the small converging angle $\alpha $ one may invoke the
Born-Oppenheimer approximation: \ $\psi \left( x,y,z\right) =\varphi \left(
x\right) \chi \left( y,z;x\right) $, in which the crossed derivatives $\frac{%
\partial }{\partial x}\chi \left( y,z;x\right) ,\frac{\partial ^{2}}{%
\partial x^{2}}\chi \left( y,z;x\right) $ are neglected, and the wave
equation takes the approximate form:

\[
-\frac{1}{\chi \left( y,z;x\right) }\left( \frac{\partial ^{2}}{\partial
y^{2}}+\frac{\partial ^{2}}{\partial z^{2}}\right) \chi \left( y,z;x\right) -%
\frac{1}{\varphi \left( x\right) }\frac{\partial ^{2}}{\partial x^{2}}%
\varphi \left( x\right) =\widetilde{\varepsilon }^{2} 
\]%
subject to the boundary conditions, Eq.(\ref{BoundCond}). \ Solutions for
the "slow" motion wave equation satisfying these boundary conditions are: $%
\chi _{n_{y},n_{z}}\left( y,z;x\right) =\frac{1}{l\left( x\right) }\sin %
\left[ q_{y}\left( x\right) \left( y-l\left( x\right) \right) \right] \sin %
\left[ q_{z}\left( x\right) \left( z+b-l\left( x\right) \right) \right] $
where: $q_{y,z}\left( x\right) =\pi n_{y,z}/l\left( x\right) ,$ $%
n_{y,z}=1,2,...$. \ 

\ \ The resulting equation for the "fast" motion is: 
\begin{eqnarray}
\left[ -\frac{d^{2}}{dx^{2}}+v_{\overrightarrow{n}}^{2}\left( x\right) %
\right] \varphi \left( x\right) &=&\widetilde{\varepsilon }^{2}\varphi
\left( x\right) \text{ , \ \ }  \TCItag{A2}  \label{FastMot} \\
\text{where}\text{: \ } &&v_{\overrightarrow{n}}^{2}\left( x\right)
=q_{y}^{2}\left( x\right) +q_{z}^{2}\left( x\right)  \nonumber
\end{eqnarray}

Thus, to first order in perturbation theory with respect to $v_{%
\overrightarrow{n}}^{2}\left( x\right) $, the energy eigenvalues of an
electron 'trapped' by the EM lenses inside the conic beam region are given
by: $\ \ $

\begin{eqnarray}
\varepsilon _{p_{x},n}^{2} &=&\hslash ^{2}c^{2}\widetilde{\varepsilon }%
^{2}+m_{0}^{2}c^{4}  \TCItag{A3} \\
&\approx &p_{x}^{2}c^{2}+m_{0}^{2}c^{4}+c^{2}\hbar ^{2}\left( \pi n/l\right)
^{2},  \nonumber \\
\ \ \ n^{2} &=&n_{y}^{2}+n_{z}^{2}.  \nonumber
\end{eqnarray}%
where \ $p_{x}=\hbar q_{x}$ is the e-beam main (longitudinal) momentum, and $%
p_{y,z}=\hbar \pi n_{y,z}/l$ , with: $n_{y,z}=1,2,...,l=\sqrt{l_{0}\left(
l_{0}+\alpha L\right) }$, its transverse momentum components in the free
propagation zone outside the EM focusing domain.

This model of the e-beam is, of course, a drastic simplification of the
actual focused beam in STEM. In particular the ideally reflecting boundary
conditions, Eq.(\ref{BoundCond}), can not be strictly realized under the
smoothly varying field generated in space by the focusing EM lenses. \ The
results of our analysis here are not expected to be very sensitive to the
fine details of the momentum distribution of the beam. We may take advantage
of that by eliminating the specific dependence of the transverse wave
numbers on the average beam radius $l$ , and replace $\left( \pi /l\right)
\left( n_{y},n_{z}\right) $ with the general symbol $\overrightarrow{q}_{tr}$%
, such that the specialized set of eigenfunctions, $\chi
_{n_{y},n_{z}}\left( y,z;x\right) $, may be replaced by a more general set $%
\chi _{\overrightarrow{q}_{tr}}\left( y,z;x\right) $.

The relativistic asymptotic (initial and final) energies of an electron
'trapped' within the beam double-cone boundary are: $\varepsilon
_{i,f}^{2}=m_{0}^{2}c^{4}+\left( p_{x}^{2}c^{2}+\frac{\hbar ^{2}\pi
^{2}n^{2}c^{2}}{l^{2}}\right) _{i,f}=\left[ m_{0}^{2}c^{4}+\left(
p_{x}^{i,f}\right) ^{2}c^{2}\right] +\left[ \left( p_{y}^{i,f}\right)
^{2}+\left( p_{z}^{i,f}\right) ^{2}\right] c^{2}$, where: $\ \left(
p_{x}^{i,f}\right) ^{2}=\left( \varepsilon _{i,f}^{2}-m_{0}^{2}c^{4}\right)
/c^{2}-\left[ \left( p_{y}^{i,f}\right) ^{2}+\left( p_{z}^{i,f}\right) ^{2}%
\right] $. \ \ 

The corresponding longitudinal momentum transfer is calculated from:

$\left( q_{x}^{i}-q_{x}^{f}\right) \left( q_{x}^{i}+q_{x}^{f}\right) =\left(
\varepsilon _{i}-\varepsilon _{f}\right) \left( \varepsilon _{i}+\varepsilon
_{f}\right) /\left( \hbar c\right) ^{2}+\left[ \left( q_{tr}^{f}\right)
^{2}-\left( q_{tr}^{i}\right) ^{2}\right] $ , where:

$\left( q_{x}^{i}+q_{x}^{f}\right) \approx 2mv/\hbar $ \ , \ $\left(
\varepsilon _{i}+\varepsilon _{f}\right) \approx \left( 2mc^{2}\right) $, so
that:

$\left( q_{x}^{i}-q_{x}^{f}\right) 2mv/\hbar \approx \left( \varepsilon
_{i}-\varepsilon _{f}\right) \left( 2mc^{2}\right) /\left( \hbar c\right)
^{2}+\left[ \left( q_{tr}^{f}\right) ^{2}-\left( q_{tr}^{i}\right) ^{2}%
\right] $ , namely:

\[
\Delta q_{x}\approx \left( \omega /v\right) +\hbar \left[ \left(
q_{tr}^{f}\right) ^{2}-\left( q_{tr}^{i}\right) ^{2}\right] /2mv 
\]

\section{Appendix B}

In this appendix, following the method developed in Ref.\cite{Maniv82}, we
consider the dielectric loss function $\func{Im}\left[ \frac{r^{f,i}\left(
k_{x},k_{y},\omega \right) }{K^{\star }}e^{-2K^{\star }b}\right] $ appearing
in Eq.(\ref{RateProbab}), and show that it is proportional to $\func{Re}%
E_{x}\left( \overrightarrow{K};-b;\omega \right) $- the electric field
component along the e-beam axis at the beam position $z=-b$. The latter is
the key ingredient in the calculation of the power loss function in the
classical limit. We shall also show in this appendix that in the long
wavelengths limit discussed in Ref.\cite{Maniv82} the dielectric loss
function reduces to the well known expression derived in Ref.\cite{Wang96}.
\ 

Our analysis starts from the expectation value of the four-vector potential (%
$A_{\nu }=\left( \varphi ,-\overrightarrow{A}\right) $ ), given by:

\begin{equation}
A_{\nu }\left( \overrightarrow{r},t\right) =\frac{1}{c}\dsum\limits_{\mu
=0}^{3}\int_{-\infty }^{\infty }dt^{\prime }\int d^{3}r^{\prime }D_{\nu ,\mu
}\left( \overrightarrow{r},\overrightarrow{r}^{\prime };t-t^{\prime }\right)
j^{ext,\mu }\left( \overrightarrow{r}^{\prime },t^{\prime }\right)  \tag{B1}
\label{Avs.Jex}
\end{equation}%
where $j^{ext,\mu }\left( \overrightarrow{r}^{\prime },t^{\prime }\right) $
is the external four-current density generated by the e-beam (with the
components $j^{ext,\nu }=\left( c\rho ^{ext},\overrightarrow{j}^{ext}\right) 
$), and $D_{\nu ,\mu }\left( \overrightarrow{r},\overrightarrow{r}^{\prime
};t-t^{\prime }\right) $ is the \textquotedblright
dressed\textquotedblright\ retarded photon Green's function, defined by the
correlator: 
\begin{equation}
D_{\nu ,\mu }\left( \overrightarrow{r},\overrightarrow{r}^{\prime
};t-t^{\prime }\right) =-i\left\langle \left[ \widehat{A}_{\nu }\left( 
\overrightarrow{r},t\right) ,\widehat{A}_{\mu }\left( \overrightarrow{r}%
^{\prime },t^{\prime }\right) \right] \right\rangle \theta \left(
t-t^{\prime }\right)  \tag{B2}  \label{DnumuDef.}
\end{equation}

The \textquotedblright bare\textquotedblright\ four-vector potential is
given by: 
\begin{eqnarray*}
A_{\nu }^{\left( 0\right) }\left( \overrightarrow{r},t\right) &=&\frac{1}{c}%
\int_{-\infty }^{\infty }dt^{\prime }\int d^{3}r^{\prime }D_{\nu }^{\left(
0\right) }\left( \overrightarrow{r}-\overrightarrow{r}^{\prime };t-t^{\prime
}\right) \\
&&\times j^{ext,\nu }\left( \overrightarrow{r}^{\prime },t^{\prime }\right)
\end{eqnarray*}%
where $D_{\nu ,\mu }^{\left( 0\right) }\left( \overrightarrow{r},%
\overrightarrow{r}^{\prime };t-t^{\prime }\right) \equiv D_{\nu }^{\left(
0\right) }\left( \overrightarrow{r}-\overrightarrow{r}^{\prime };t-t^{\prime
}\right) \delta _{\nu ,\mu }$ , is the \textquotedblright
bare\textquotedblright\ retarded photon propagator in the Lorentz gauge,
which is given by: $D_{\nu }^{\left( 0\right) }\left( \overrightarrow{r}-%
\overrightarrow{r}^{\prime };t-t^{\prime }\right) =\eta _{\nu }\theta \left(
t-t^{\prime }\right) \delta \left( \left\vert \overrightarrow{r}-%
\overrightarrow{r}^{\prime }\right\vert /c+t-t^{\prime }\right) /\left\vert 
\overrightarrow{r}-\overrightarrow{r}^{\prime }\right\vert $, with: $\eta
_{\nu }=\left( 
\begin{array}{c}
1\text{ , \ }\nu =0 \\ 
-1\text{ , }\nu =1,2,3%
\end{array}%
\right) $.

The corresponding Fourier transforms with respect to the spatial coordinates
parallel to the surface, with wave vector $\overrightarrow{K}=\left(
k_{x},k_{y}\right) $ are:

$A_{\nu }\left( \overrightarrow{K};z;\omega \right) =\frac{1}{c}%
\dsum\limits_{\mu =0}^{3}\int dz^{\prime }D_{\nu ,\mu }\left( 
\overrightarrow{K};z,z^{\prime };\omega \right) $

$\times j^{ext,\mu }\left( \overrightarrow{K};z^{\prime };\omega \right) $,
and: \ $D_{\nu }^{\left( 0\right) }\left( \overrightarrow{K},z,z^{\prime
};\omega \right) =D_{\nu }^{\left( 0\right) }\left( K^{\star },\left\vert
z-z^{\prime }\right\vert \right) =\frac{\eta _{\nu }}{2\pi K^{\star }}%
e^{-K^{\ast }\left\vert z-z^{\prime }\right\vert }$. \ 

Our explicit expression for the external 4-current density associated with
the e-beam is: 
\begin{eqnarray}
&&j^{ext,\mu }\left( \overrightarrow{K};z^{\prime };\omega \right) 
\TCItag{B3}  \label{jex} \\
&=&\left( 
\begin{array}{c}
-ce\delta \left( z+b\right) \delta \left( vk_{x}-\omega \right) \text{ \ \ \
\ \ \ \ },\text{ \ \ }\mu =0 \\ 
-e\left( \frac{\omega }{k_{x}}\right) \delta \left( z+b\right) \delta \left(
vk_{x}-\omega \right) ,\text{ \ \ }\mu =1 \\ 
\text{ }0\text{ \ \ \ \ \ \ \ \ \ \ \ \ \ \ \ \ \ \ \ \ \ \ \ \ \ \ \ \ \ \
\ \ \ \ \ \ \ \ }\ \ \ ,\text{ }\mu =2,3%
\end{array}%
\right)  \nonumber
\end{eqnarray}

For $z,z^{\prime }\leq 0$ \ , i.e. both on the vacuum side of the dielectric
slab, occupying the space: $2c^{\star }>z>0$ , the lateral Fourier transform
of Eq.(\ref{DnumuDef.}) can be written in the form: 
\begin{equation}
D_{\nu ,\mu }\left( \overrightarrow{K},z,z^{\prime };\omega \right) =\frac{%
\eta _{\nu }}{2\pi K^{\star }}\left[ 
\begin{array}{c}
\delta _{\nu ,\mu }e^{-K^{\ast }\left\vert z-z^{\prime }\right\vert } \\ 
-\left( r_{\nu ,\mu }^{\left( odd\right) }+r_{\nu ,\mu }^{\left( even\right)
}\right) e^{K^{\star }\left( z+z^{\prime }\right) }%
\end{array}%
\right]  \tag{B4}  \label{Dnumu}
\end{equation}%
where the generalized reflection four-matrices for incident waves, which are
either symmetric or antisymmetric with respect to the slab center, are given
respectively by (see Ref.\cite{Maniv82}):

\begin{eqnarray}
\mathbf{r}^{\left( odd,even\right) } &\mathbf{=}&\frac{1}{2}\left( \mathbf{W}%
^{\left( odd,even\right) }-\mathbf{I}\right) ,\text{ }  \TCItag{B5}
\label{rmat} \\
\text{\ \ }\mathbf{W}^{\left( odd,even\right) } &\mathbf{\equiv }&\left( 
\mathbf{U}^{\left( odd,even\right) }\mathbf{+I}\right) ^{-1}  \nonumber
\end{eqnarray}%
The definition of the matrix $\mathbf{U}$ can be found in Ref.\cite{Maniv82}%
).

In the limit of a semi-infinite slab ($c^{\star }\rightarrow \infty $), \ $%
\mathbf{r}^{\left( odd\right) }=\mathbf{r}^{\left( even\right) }\equiv 
\mathbf{r}$ , so that:

\[
r_{\nu ,\mu }^{\left( odd\right) }+r_{\nu ,\mu }^{\left( even\right)
}\rightarrow 2r_{\nu ,\mu }\text{ \ , \ \ }c^{\star }\rightarrow \infty 
\]

and:\ \ \ 

\begin{eqnarray*}
A_{\nu }\left( \overrightarrow{K};z;\omega \right) &=&-\frac{1}{c}\frac{\eta
_{\nu }}{2\pi K^{\ast }}\left( \delta _{\nu ,0}e^{-K^{\star }\left\vert
z+b\right\vert }-2r_{\nu ,0}e^{K^{\star }\left( z-b\right) }\right) \\
&&\times ce\delta \left( vk_{x}-\omega \right) \\
&&-\frac{1}{c}\frac{\eta _{\nu }}{2\pi K^{\ast }}\left( \delta _{\nu
,1}e^{-K^{\star }\left\vert z+b\right\vert }-2r_{\nu ,1}e^{K^{\star }\left(
z-b\right) }\right) \\
&&\times e\left( \frac{\omega }{k_{x}}\right) \delta \left( vk_{x}-\omega
\right)
\end{eqnarray*}

where at $z=-b$: \ 
\begin{eqnarray}
A_{\nu }\left( \overrightarrow{K};-b;\omega \right) &=&-\frac{e}{2\pi
K^{\star }}\delta \left( vk_{x}-\omega \right) \eta _{\nu }  \TCItag{B6}
\label{Anu(K)} \\
&&\left\{ 
\begin{array}{c}
\left[ \delta _{\nu ,0}+\left( \frac{\omega }{ck_{x}}\right) \delta _{\nu ,1}%
\right] \\ 
-\left[ 2r_{\nu ,0}+\left( \frac{\omega }{ck_{x}}\right) 2r_{\nu ,1}\right]
e^{-2K^{\star }b}%
\end{array}%
\right\}  \nonumber
\end{eqnarray}

To simplify the calculation we shall consider in what follows the special
case when the wave vector $\overrightarrow{K}$ is parallel to the e-beam
direction, which was selected along the $x$-axis, so that $k_{y}=0$ , and $%
K=k_{x}$.

Thus, the electric field component along the e-beam axis at the beam
position $z=-b$ , $E_{x}\left( \overrightarrow{K};-b;\omega \right) $ , can
be calculated from the explicit expressions for the potentials in Lorentz
gauge, namely:

$E_{x}\left( \overrightarrow{K};-b;\omega \right) =-ik_{x}A_{0}\left( 
\overrightarrow{K};-b;\omega \right) -i\left( \omega /c\right) A_{1}\left( 
\overrightarrow{K};-b;\omega \right) $

$=\left[ 
\begin{array}{c}
-ik_{x}\left( r_{0,0}+\left( \frac{\omega }{ck_{x}}\right) r_{0,1}\right) -
\\ 
i\left( \omega /c\right) \left( r_{1,0}+\left( \frac{\omega }{ck_{x}}\right)
r_{1,1}\right)%
\end{array}%
\right] e^{-2K^{\star }b}=$

$-ik_{x}\left[ r_{0,0}+\left( \frac{\omega }{ck_{x}}\right) r_{0,1}+\left( 
\frac{\omega }{ck_{x}}\right) r_{1,0}+\left( \frac{\omega }{ck_{x}}\right)
^{2}r_{1,1}\right] e^{-2K^{\star }b}$, or (by exploiting the symmetry
property: $r_{1,0}=-r_{0,1}$\cite{Maniv82}):

\begin{equation}
E_{x}\left( \overrightarrow{K};-b;\omega \right) =-ik_{x}\left[
r_{0,0}+\left( \frac{\omega }{ck_{x}}\right) ^{2}r_{1,1}\right]
e^{-2K^{\star }b}  \tag{B7}  \label{E_x}
\end{equation}

This expression should be compared to the dielectric response function: 
\begin{eqnarray*}
&&r^{f,i}\left( k_{x},k_{y},\omega \right) \\
&=&%
\begin{array}{c}
r_{0,0}+\left( \hbar q_{x}^{i}/mc\right) r_{0,1}+\left( \hbar
q_{x}^{f}/mc\right) r_{1,0} \\ 
+\left( \hbar ^{2}q_{x}^{f}q_{x}^{i}/\left( mc\right) ^{2}\right) r_{1,1}%
\end{array}%
\end{eqnarray*}%
which may be simplified (again due to the symmetry $r_{1,0}=-r_{0,1}$ and
the inequality $\left\vert \Delta q_{x}\right\vert \approx \omega /v\ll
mc/\hbar $ ) to: 
\begin{eqnarray}
r^{f,i}\left( k_{x},k_{y},\omega \right) &=&%
\begin{array}{c}
r_{0,0}+\left( v/c\right) ^{2}r_{1,1} \\ 
+\left( \hbar \Delta q_{x}/mc\right) \left[ r_{0,1}-\left( v/c\right) r_{1,1}%
\right]%
\end{array}
\nonumber \\
&\approx &%
\begin{array}{c}
r_{0,0}+\left( v/c\right) ^{2}r_{1,1} \\ 
+\left( \hbar \omega /mcv\right) \left[ r_{0,1}-\left( v/c\right) r_{1,1}%
\right]%
\end{array}
\nonumber \\
&\approx &r_{0,0}+\left( v/c\right) ^{2}r_{1,1}  \TCItag{B8}
\label{DielectResp-final}
\end{eqnarray}

Thus, since in the "classical" limit the prefactor of $r_{1,1}$ in Eq.(\ref%
{E_x}) $\left( \frac{\omega }{ck_{x}}\right) ^{2}\longleftrightarrow \left(
v/c\right) ^{2}$, we find that: \ 
\[
\func{Re}E_{x}\left( \overrightarrow{K};-b;\omega \right) \propto \func{Im}%
\left[ \left( r_{0,0}+\left( v/c\right) ^{2}r_{1,1}\right) e^{-2K^{\star }b}%
\right] 
\]

Exploiting the continuity equation to connect various components of the
matrix $\mathbf{U}$\cite{Maniv82}:

$k_{x}U_{11}-\left( \omega /c\right) U_{01}=-K^{\ast }U_{31}$, \ and \ $%
k_{x}U_{13}-\left( \omega /c\right) U_{03}=-K^{\ast }U_{33}$, and noting the
symmetry relation: $U_{31}=-U_{13}$, it can be shown that: 
\[
r_{0,0}+\left( \frac{\omega }{ck_{x}}\right) r_{0,1}=-\frac{\left( \omega
/ck_{x}\right) U_{01}+U_{00}}{2\left( 1+Tr\mathbf{U}\right) } 
\]

Furthermore, the continuity equation also implies:

$k_{x}U_{01}+\left( \omega /c\right) U_{00}=K^{\ast }U_{30}$, \ and \ $%
k_{x}U_{13}-\left( \omega /c\right) U_{03}=-K^{\ast }U_{33}$, so that

since $U_{30}=U_{03}$, we also find that:\ \ 
\[
r_{1,0}+\left( \frac{\omega }{ck_{x}}\right) r_{1,1}=\frac{U_{01}-\left(
\omega /ck_{x}\right) U_{11}}{2\left( 1+Tr\mathbf{U}\right) } 
\]

\ Consequently: $\ r_{0,0}+\left( v/c\right) ^{2}r_{1,1}\rightarrow
r_{0,0}+\left( \frac{\omega }{ck_{x}}\right) ^{2}r_{1,1}=-\frac{%
U_{00}+\left( \omega /ck_{x}\right) ^{2}U_{11}}{2\left( 1+Tr\mathbf{U}%
\right) }$

and in the long wavelength limit, where $1+Tr\mathbf{U=}\frac{\left(
\varepsilon K^{\ast }+Q\right) }{2\varepsilon K^{\ast }}$ and $\varepsilon $
is the bulk optical (frequency dependent) dielectric function of the
platelet, we find:

\begin{equation}
U_{0,0}=\frac{\left( 1-\varepsilon \right) }{2\varepsilon }\left( 1+\frac{%
\left( \omega /c\right) ^{2}}{2K^{\ast 2}}\right)  \tag{B9}  \label{U00}
\end{equation}

\begin{equation}
U_{1,1}=\frac{\left( \varepsilon -1\right) }{\varepsilon }\frac{\left(
\omega /c\right) ^{2}}{4K^{\ast 2}}\left[ 1+\left( \varepsilon -1\right) 
\frac{K^{\ast 2}}{\left( Q+K^{\ast }\right) ^{2}}\right]  \tag{B10}
\label{U11}
\end{equation}%
so that finally:

\begin{eqnarray*}
&&\func{Re}E_{x}\left( \overrightarrow{K};-b;\omega \right) \\
&\propto &\func{Im}\left\{ \frac{e^{-2K^{\ast }b}}{K^{\ast }}\left[ 
\begin{array}{c}
\frac{K^{\ast }\left( \varepsilon -1\right) }{\left( \varepsilon K^{\ast
}+Q\right) }+ \\ 
\left( \frac{v}{c}\right) ^{2}\left( \frac{\left( K^{\ast }-Q\right) }{%
\left( Q+K^{\ast }\right) }+\frac{\left( 1-\varepsilon \right) K^{2}}{\left(
Q+K^{\ast }\right) \left( \varepsilon K^{\ast }+Q\right) }\right)%
\end{array}%
\right] \right\}
\end{eqnarray*}%
which is equivalent to surface dielectric loss function obtained in Ref.\cite%
{Wang96} by using Maxwell's equations with macroscopic boundary conditions.

\bigskip

\bigskip \textbf{Acknowledgements: }

We thank Boris Lembrikov for helpful discussions. This research was
supported by ARGENTINIAN\ RESEARCH\ FUND, and by the fund for the promotion
of research at the Technion.

. \qquad \qquad \qquad \qquad \qquad \qquad \qquad \qquad \qquad \qquad
\qquad \qquad \qquad \qquad \qquad \qquad \qquad \qquad \qquad \qquad \qquad
\qquad \qquad \qquad \qquad \qquad \qquad \qquad \qquad \qquad \qquad \qquad
\qquad \qquad \qquad \qquad \qquad \qquad \qquad \qquad \qquad \qquad \qquad
\qquad \qquad \qquad \qquad \qquad \qquad \qquad \qquad \qquad \qquad \qquad
\qquad \qquad \qquad \qquad \qquad \qquad \qquad \qquad \qquad \qquad \qquad
\qquad\

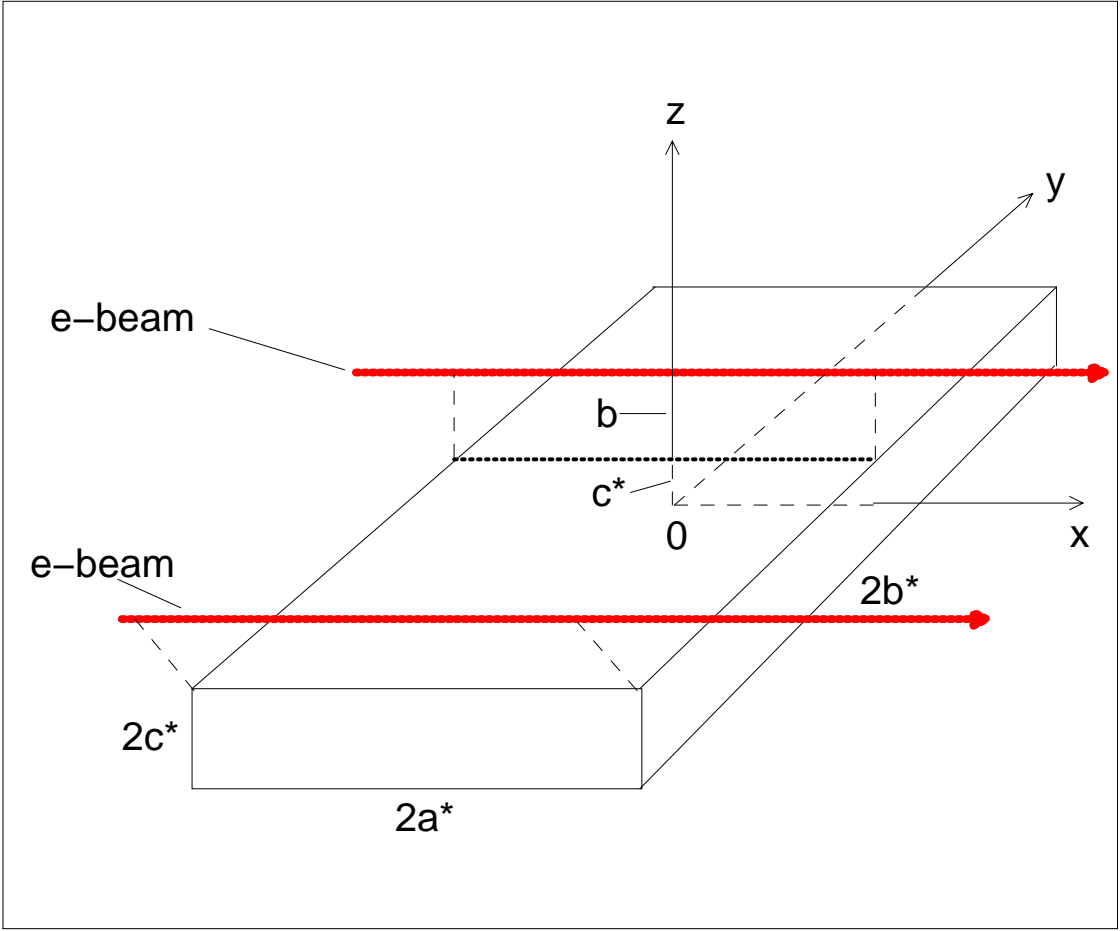
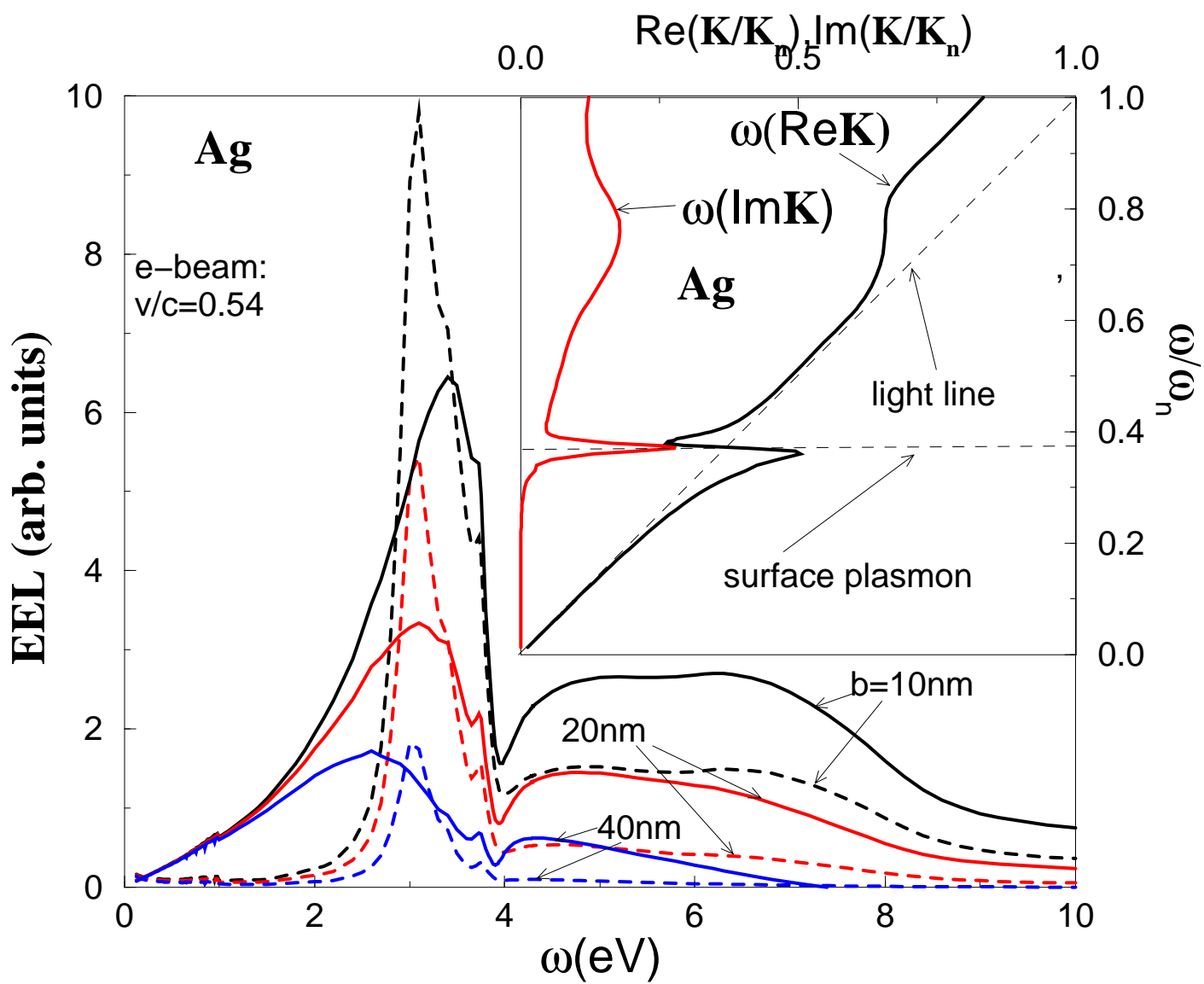
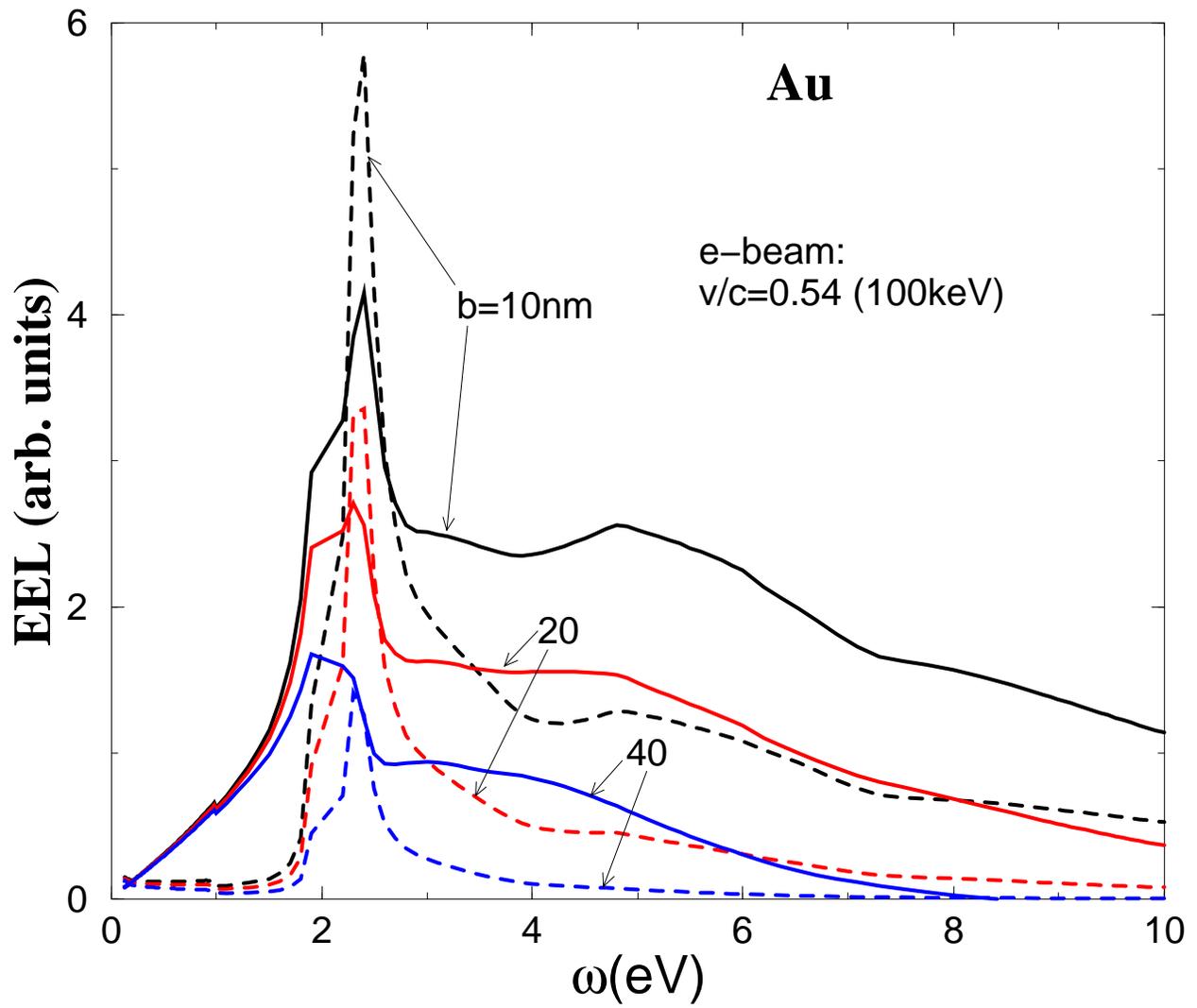
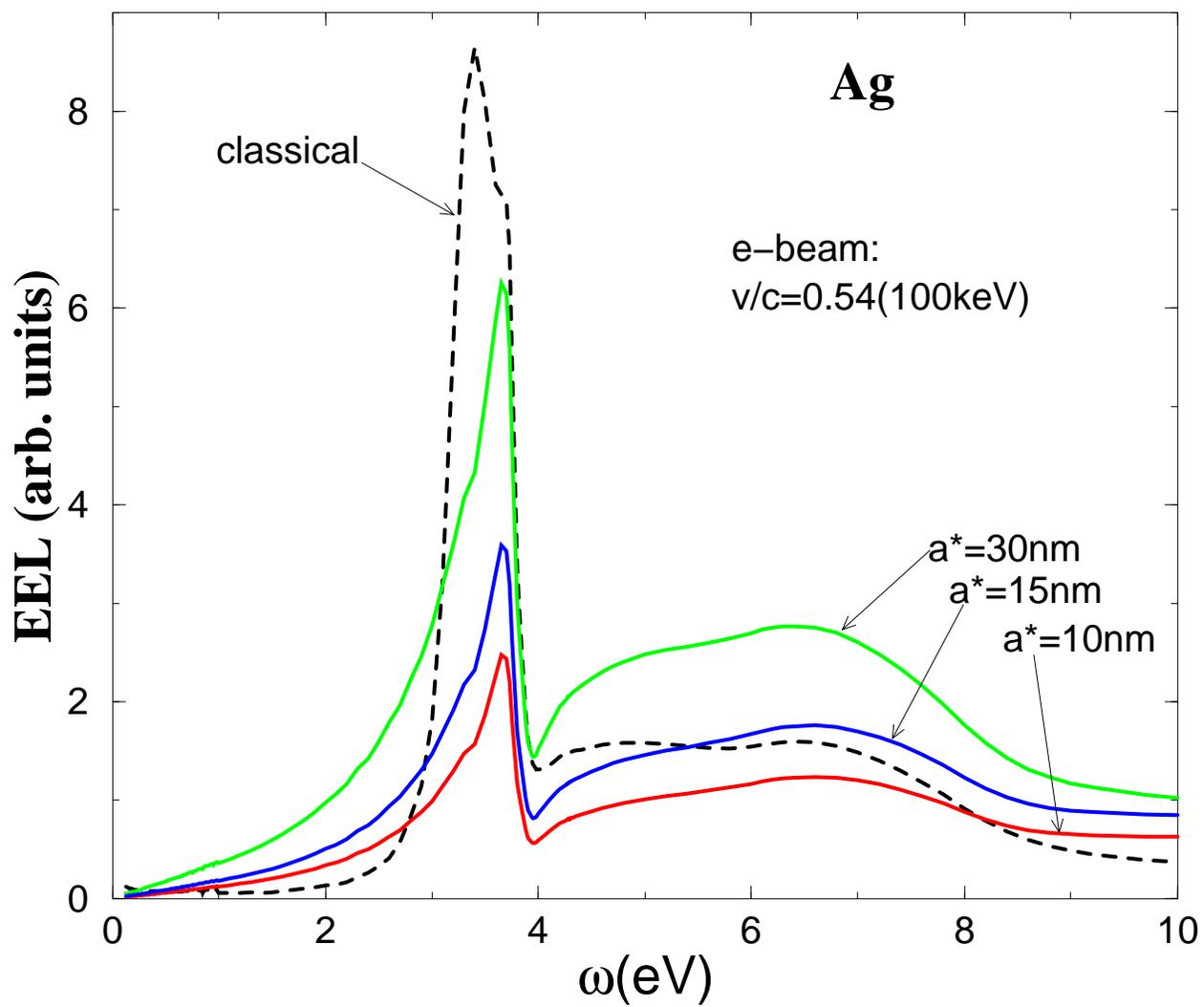
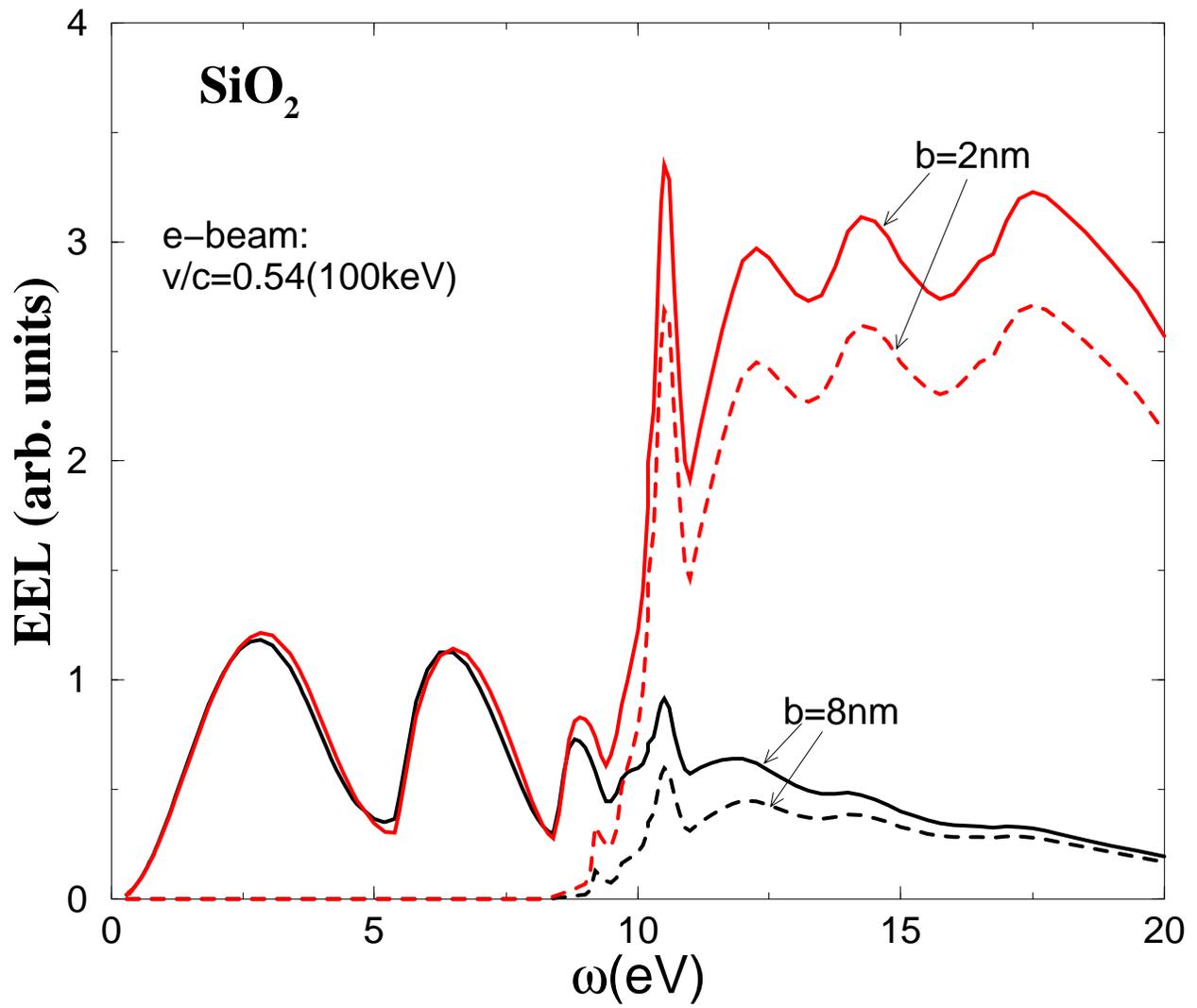
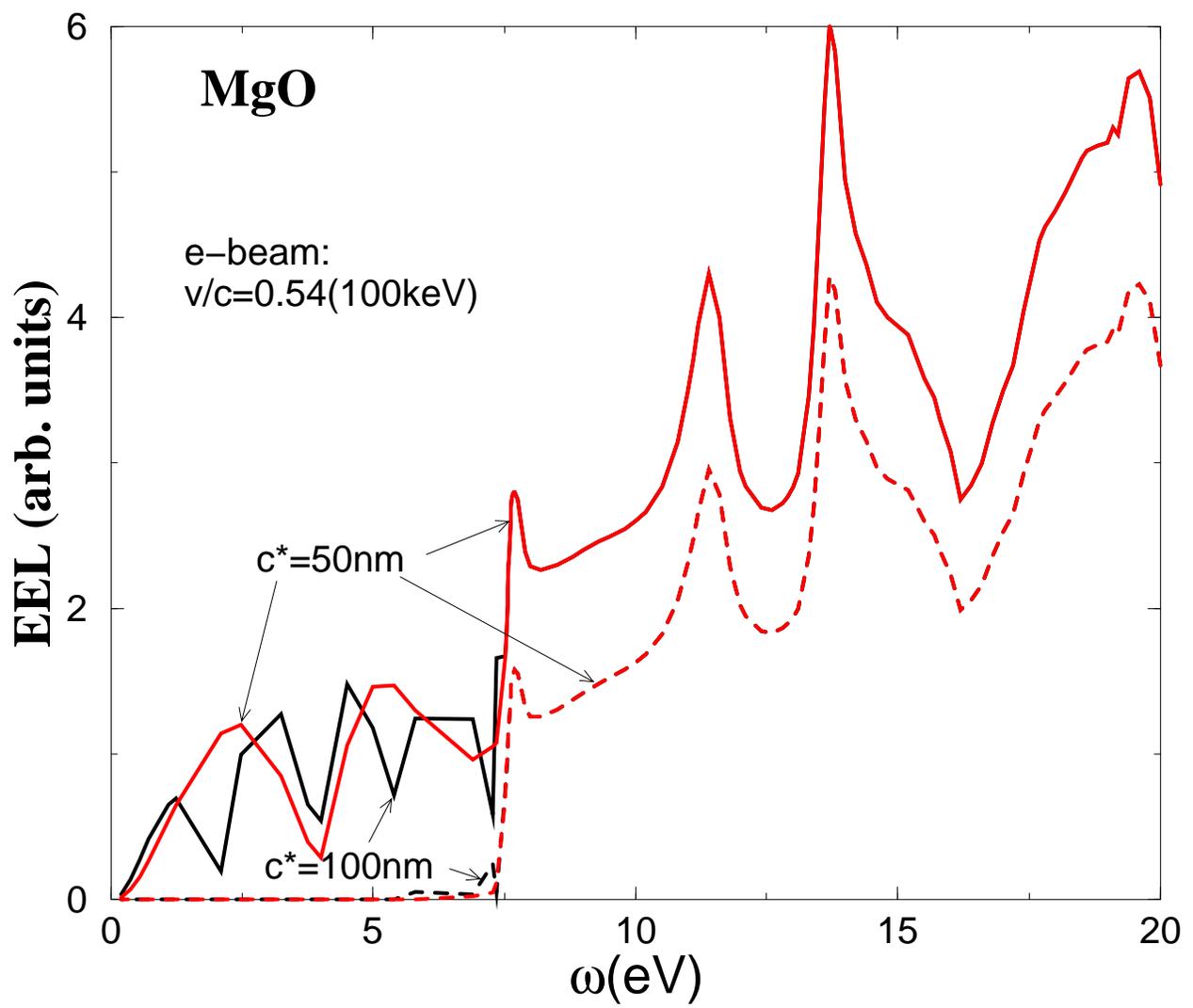
\end{document}